# Improving Reproducibility of Sputter Deposited Ferroelectric Wurtzite Al$_{0.6}$Sc$_{0.4}$N Films using In-situ Optical Emission Spectrometry

*Daniel Drury\*, Keisuke Yazawa, Allison Mis, Kevin Talley, Andriy Zakutayev, Geoff L. Brennecka\**

D. D., Dr. K. Y., A. M., Dr. G. B.

Department of Metallurgical and Materials Engineering, Colorado School of Mines, 80401, United States

D. D., Dr. K. Y., A. M., Dr. K. T., Dr. A. Z.

Materials Science Center, National Renewable Energy Laboratory, 80401, United States

E-mail: ddrury@mymail.mines.edu, geoff.brennecka@mines.edu

**Abstract**

High-Sc Al$_{1-x}$Sc$_x$N thin films are of tremendous interest because of their attractive piezoelectric and ferroelectric properties, but overall film quality and reproducibility are widely reported to suffer as $x$ increases. In this study, we correlate the structure and electrical properties of Al$_{0.6}$Sc$_{0.4}$N with in-situ observations of glow discharge optical emission during growth. This in-situ technique uses changes in the Ar(I) and N$_2$(I) emission lines of the glow discharge during growth to identify films that subsequently exhibit unacceptable structural and electrical performance. We show that a steady deposition throughout film growth produces ferroelectric Al$_{0.6}$Sc$_{0.4}$N with a reversible 80 μC cm$^{-1}$ polarization and 3.1 MV cm$^{-1}$ coercive field. In other films deposited using identical settings, fluctuations in both Ar(I) and N$_2$(I) line intensities correspond to decreased wurtzite phase purity, nm-scale changes to the film microstructure, and a non-ferroelectric response. These results illustrate the power of optical emission spectroscopy for tracking changes when fabricating process-sensitive samples such as high-Sc Al$_{1-x}$Sc$_x$N films.

1. Introduction

Interest in the growth of AlN based group-III metal nitride semiconductor alloys such as (Al,Ga)N and (Al,In,Ga)N continues to increase as the associated application space expands. First the interest was driven by their tunable properties for light emitting diodes (LEDs) and other optoelectronic applications [1] and by their wide band gap (WBG) semiconductor properties for high electron mobility transistors (HEMTs) in RF and power electronic applications.[2] Scandium additions to AlN were first reported to produce significant increases in piezoelectric response in 2009[3] and were very quickly adopted for the piezoelectric thin film devices such as film bulk acoustic resonators (FBARs) in cell phones.[4] Recent reports on ferroelectricity in Al$_{1-x}$Sc$_x$N for $x \geq 0.1$ have sparked additional scientific interest as the first wurtzite ferroelectric material[5,6] and significant technological interest as a candidate for hybrid logic-in-memory devices.

Large scale commercial applications of piezoelectric Al$_{1-x}$Sc$_x$N are still limited to relatively modest levels of Sc substitution (x<0.1), though there are many literature reports of Al$_{1-x}$Sc$_x$N films having $x \geq 0.1$ with exceptional properties.[3-12] It has been widely reported that increasing Sc content, desired for increased piezoelectric response and lower-field ferroelectric switching, increases the challenge of growing high quality textured Al$_{1-x}$Sc$_x$N films[13-15]. This challenge is driven in no small part by the thermodynamic driving force for phase separation in this system.[16,17] This degradation in film properties for large $x$ values has been correlated with wurtzite-(0002) texture reduction with the presence of misoriented grains[11] and the associated difficulties with reliably controlling the stress state in sputtered films.[18] However, in each case the conclusions are correlational and are drawn largely or entirely from ex-situ measurements, so they are of limited value for deposition process control.

Many in-situ monitoring techniques have been reported to tune the deposition of AlN thin films: glow discharge optical emission spectrometry (GD-OES),[19,20] reflection high-energy electron diffraction (RHEED),[21,22] reflectance interferometry (RI),[23] and ellipsometry.[24] While useful as surface diagnostic tools for understanding how the film develops, RHEED, RI, and ellipsometry are costly methods which require significant adaptation to the system such as beam alignment. On the other hand, GD-OES only requires line of sight to the glow discharge. GD-OES is also able to evaluate the target poison mode for reactive depositions, which is an important step for optimizing rf sputtered nitrides.[20] However, there presently lacks literature on incorporating GD-OES when growing $Al_{1-x}Sc_xN$, which is an opportunity considering the difficulty of producing a purely wurtzite phase, particularly for large values of $x$.

In this paper, we investigate the process of sputtering $Al_{1-x}Sc_xN$ thin films using GD-OES with support from residual gas analysis (RGA), which are two well-known, non-intrusive, in-situ monitoring techniques. Consistent with the increased variability of film quality previously reported, $Al_{0.6}Sc_{0.4}N$ films deposited in this study under identical settings for preconditioning steps and growth parameters exhibit drastically different structures and electrical properties. While the precise mechanism(s) for this disparity remain(s) unclear, we clearly correlate changes in phase content and microstructure with deposition process fluctuations observed by time-resolved GD-OES. These process fluctuations have critical effects on resulting electrical properties, exemplified by the clear ferroelectric switching of single-phase homogeneous $Al_{0.6}Sc_{0.4}N$ deposited from a continuous process and the absence of ferroelectric polarization reversal in two-phase $Al_{0.6}Sc_{0.4}N$ that resulted from observed process fluctuations of unknown origin.

2. Results and discussion

**Figure 1a** shows the XRD *θ-2θ* patterns for films deposited using the same deposition conditions and pre-treatments that have previously been optimized for this chamber (see **Experimental Section**). The patterns indicate that one film has a pure c-axis textured wurtzite structure (wz-AlScN) and the other is a mixed wurtzite/rocksalt structure (wz/rs-AlScN). The (0002) peak appears at 36.3° and corresponds to a c-axis lattice parameter of 0.494 nm for each sample which closely match previous reports for $Al_{0.6}Sc_{0.4}N$.[11] Additionally, there is a significant phase difference as indicated by the appearance of the (111) rocksalt peak at 35.7° as previously reported[3,25] and in line with theoretical work on the thermodynamic relationship between wurtzite and rocksalt in $Al_{1-x}Sc_xN$.[17] For samples grown using the same deposition conditions, it is interesting that there is a large discrepancy in phases. For both samples, the wz-(10-10) peak appears at 31.5° in Figure 1a. **Figure 1b** shows the (0002) peak profile in *χ* space. This provides a similar film structure property as an *ω*-rocking curve which measures the out of plane texture of a particular *2θ* position. 2-dimensional XRD patterns for each sample are provided in **Figure S1**. In Figure 1b we see similar out of plane texture between the two films, albeit a slightly lower FWHM for wz-AlScN at 2.8° than the wz/rs-AlScN sample with 3.1°. Based off the broad *χ* FWHM (>10°), the (10-10) peak has poor out of plane texture.

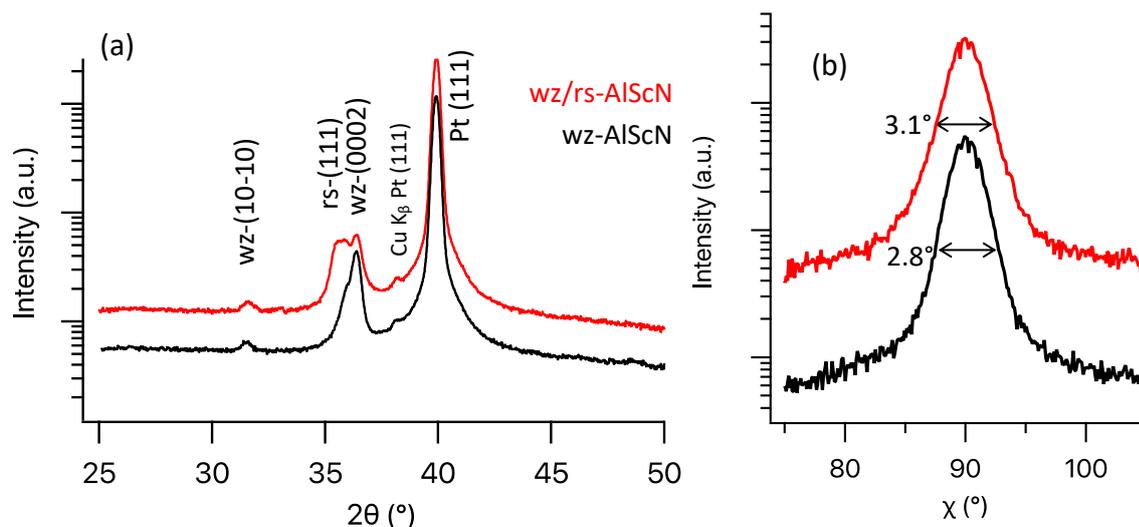

**Figure 1** 1D XRD patterns of wz-Al$_{0.6}$Sc$_{0.4}$N and wz/rs-Al$_{0.6}$Sc$_{0.4}$N on platinized silicon substrates integrated from 2D XRD detector. (a) *2θ* patterns between 25°-50° and (b) *χ* pattern between 75°-105° with associated FWHM values.

Bright field scanning transmission electron microscopy (BF-STEM) images in **Figure 2** revealed a columnar structure in both films with thicknesses of 325 nm and 337 nm for wz-AlScN and wz/rs-AlScN, respectively. The wz-AlScN sample exhibits a uniform structure through the thickness of the film, whereas striations are apparent in the wz/rs-AlScN sample as indicated by the timestamps. Assuming an overall constant deposition rate, the features occur at minutes 44, 56, 59, and 74 of the 110 minute deposition. Selected area electron diffraction (SAED) patterns of the film bulk in Figure 2b,d reveal that the films grow along the [0001] direction. The lack of (10-10) reflections in Figure 2b is a matter of sample alignment in the TEM. TEM-EDS maps and profiles show no indication of chemical segregation in either sample (**Figure S2**).

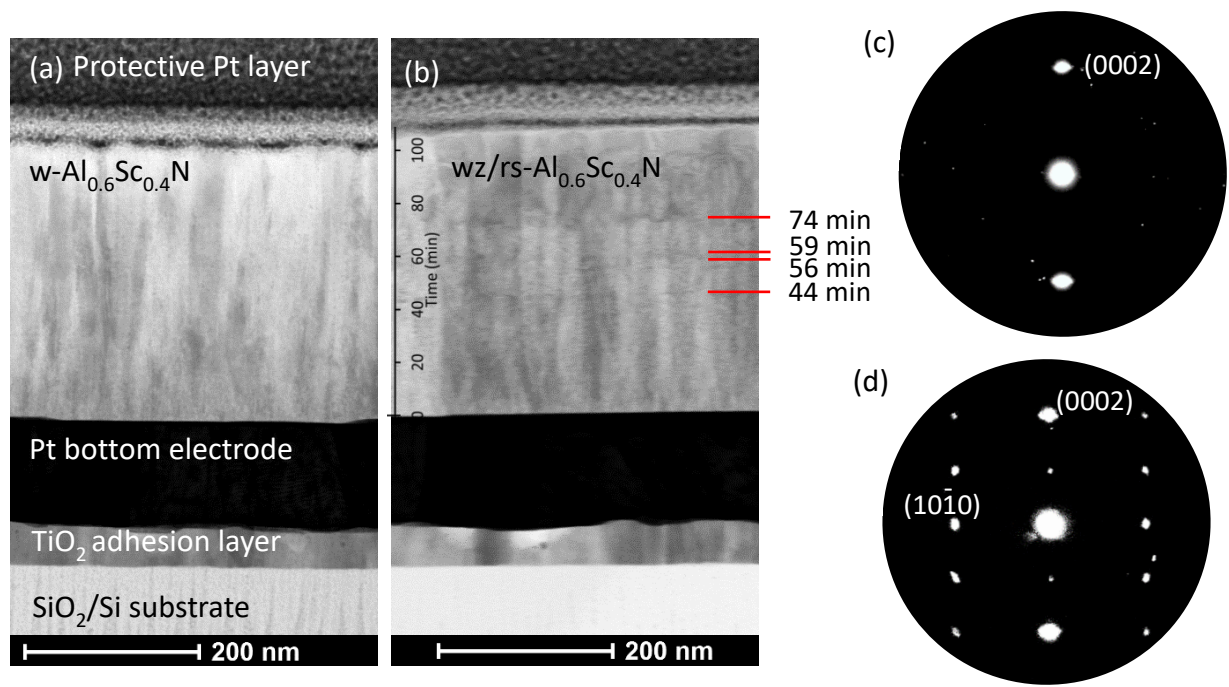

**Figure 2.** Bright field STEM images of wz-Al$_{0.6}$Sc$_{0.4}$N in (a) shows a uniform microstructure in the cross section, while for wz/rs-Al$_{0.6}$Sc$_{0.4}$N in (b) it exhibits striations as marked by the timestamps. Resultant SAED patterns for (c) wz-AlScN and (d) wz/rs-AlScN showing out of plane (0002) film texture.

Optical emission of the glow discharge near the substrate surface was collected for all growths. Lines at 337.0 nm and 696.5 nm which respectively correspond to excited molecular N$_2$(I) and atomic Ar(I) states were tracked during the deposition (**Figure 3a**). A representative full emission spectrum is provided in **Figure S3**. There are four peaks in both the Ar and N$_2$ lines for the wz/rs-AlScN sample that occur at times 44, 55, 59, and 71 minutes. These peaks align in time with the microstructure features in Figure 2c. However, the nearly uniform emission lines for wz-AlScN points toward a stable deposition which led to a single phase and pure (0002) textured film. The variation in the intensity of the lines indicates a change in glow discharge emission. While there could be multiple sources for this change (change in target poisoning and/or forward power, target flaking, magnetron configuration, etc.), the effects on the film structure and properties are clear and unambiguous.

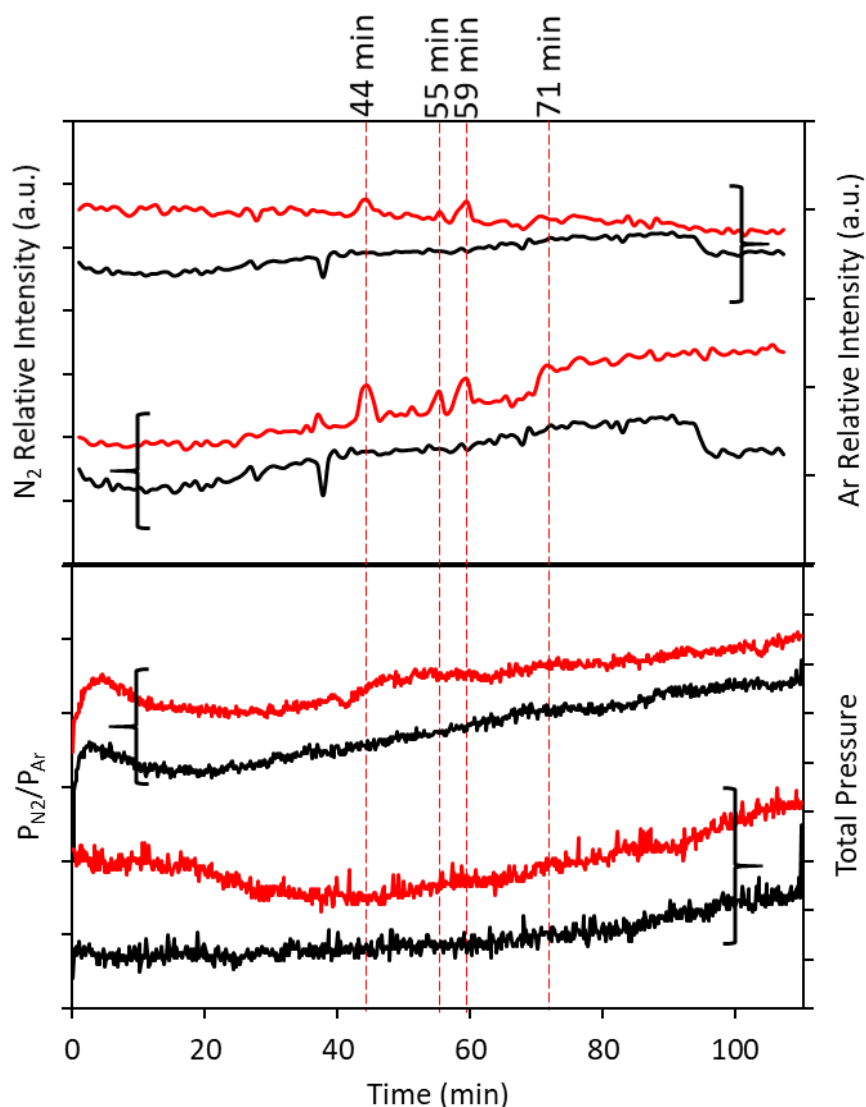

**Figure 3.** In-situ data collected by OES and RGA. (a) Relative intensities of $N_2(I)$ and $Ar(I)$ for both the wz-AlScN (black) and wz/rs-AlScN (red) samples. (b) Partial pressure ratio of $N_2$/Ar and total pressure throughout the deposition.

**Figure 3b** compares the ratio of the partial pressures of $N_2$/Ar collected by the RGA during the deposition. The lack of clear features (i.e., peaks or troughs) in the ratio scans indicates a stable pressure and gas flow rate during both film growths. This demonstrates that changes in GD-OES line intensities are separate from either a change in gas ratio or total pressure. The gradual slope of both emission lines is related to the slight changes in total pressure throughout the deposition for both samples.

Polarization-electric field (P-E) hysteresis loops were measured on parallel plate capacitors fabricated from both thin film samples. In **Figure 4a** the wz-AlScN PE loop shows a saturated hysteresis loop indicative of the highly c-axis textured structure of the film. With a remanent polarization of 80 μC cm$^{-1}$ and a coercive field of 3.1 MV cm$^{-1}$, the properties of this sample are consistent with previous reports on $x$=0.4 chemistries in $Al_{1-x}Sc_xN$.[5,6] There is a 7.5% error associated with the polarization values due to uncertainty of the device area. In comparison, the mixed phase layered wz/rs-AlScN sample did not exhibit ferroelectric switching up to 4.1 MV cm$^{-1}$ (the maximum electric field sustained before catastrophic device failure) which is related to the presence of the rocksalt phase. Thus, the lack of ferroelectric properties in wz/rs-AlScN can be correlated with observations from the GD-OES during film growth.

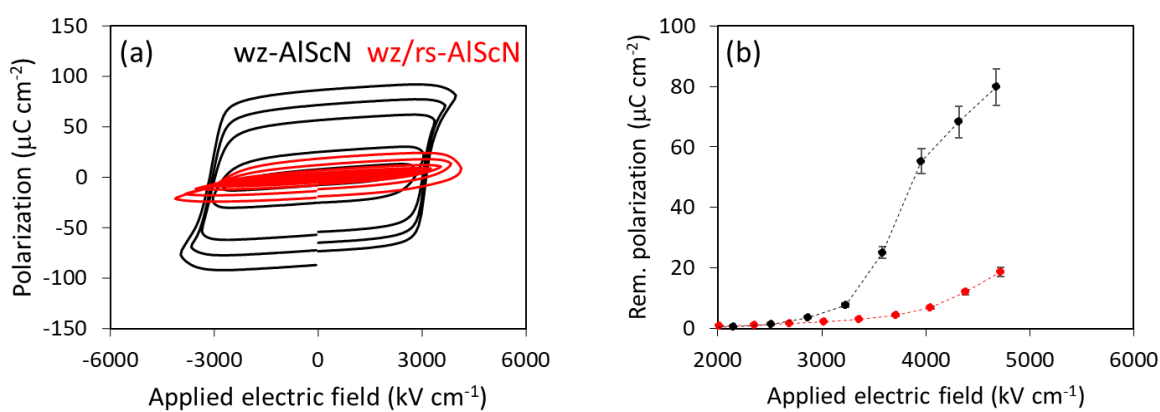

**Figure 4.** P-E loop hysteresis measurements of (a) wz-$Al_{0.6}Sc_{0.4}N$ and (b) wz/rs-$Al_{0.6}Sc_{0.4}N$ using a 10kHz excitation frequency. (b) Remanent polarization for both samples.

## 3. Conclusion

In summary, in-situ time-resolved GD-OES has been correlated with changes in Al$_{0.6}$Sc$_{0.4}$N film crystal structure and microstructure when using the same growth parameters. These changes apparent in the XRD and BF-STEM measurements reveal that phase segregation into rocksalt and wurtzite originates as a fluctuation in the glow discharge chemistry. The in-situ GD-OES technique also allows for predicting film properties since the peaks in the Ar(I) and N$_2$(I) emission lines relate to changes in the glow discharge which directly influences growth conditions. The results show that a steady deposition without fluctuations throughout film growth produces ferroelectric wz-Al$_{0.6}$Sc$_{0.4}$N with a reversible 80 µC cm$^{-1}$ polarization and 3.1 MV cm$^{-1}$ coercive field. Monitoring the deposition will help to mitigate rocksalt phase formation and improve process reliability for emerging ferroelectric Al$_{1-x}$Sc$_x$N especially when pushing toward higher $x$.

## 4. Experimental Section/Methods

Thin-film Al$_{0.6}$Sc$_{0.4}$N samples were produced via reactive radio-frequency (rf) magnetron sputtering on (111) textured platinized silicon substrates (Pt/TiO$_2$/SiO$_2$/Si) with a rocking curve (ω) FWHM <1.3°. The growth chamber had a base pressure of 5x10$^{-7}$ Torr. During the deposition, the pressure was set to 2 mTorr with gas flow rates of 15/5 sccm for Ar/N$_2$. The substrate temperature was set to 400 °C while rotating at 20 rpm at a distance of 16.5 cm from the target. A 3" Al$_{0.6}$Sc$_{0.4}$ alloy target (Stanford Advanced Materials, 99.9 at. % pure) was used with a 6.6 W cm$^{-1}$ forward power density. The target was pre-sputtered for 60 minutes under the same parameters as the deposition to reduce target surface oxidation.

In-situ monitoring consisted of an OES and an RGA. The light emitted from the glow discharge was collected through a 3° collimator lens aimed at the substrate surface by an EPP2000-UVN-SR spectrometer (StellarNet Inc) with a 0.5 nm resolution (**Figure S4**). The fiber optic feedthrough was mounted by a CF flange to the deposition chamber. The relative intensity of the Ar(I) and N$_2$(I) lines as a function of time $t$ plotted in Figure 3a is a ratio of the *I(t)/I$_o$* after subtracting the background signal which was collected prior to striking the glow discharge, where *I$_o$* is the initial spectrum collect as a reference. For OES collection, 10 spectra were subsequently collected for 5 seconds each and then averaged, which led to a 50 second time interval between each data point. This was performed to increase the signal to noise since the glow discharge density is much lower near the substrate surface than the target surface.

To investigate the crystal structure, X-ray diffraction (Bruker D8 Discover) and selected area electron diffraction (SAED) were employed. To produce 1-dimensional plots (Intensity vs. *2θ*) from the 2-dimensional detector on the Bruker XRD which simultaneously collects intensity in *χ* and *2θ* space during a scan, the intensity is integrated across *χ* (60°-120°). Scanning Transmission Electron Microscope (STEM) images were acquired using an FEI Talos F200X at 200 keV with a camera length of 77 mm. Cross-sectional lamellae were prepared using an FEI Helios Nanolab 600i FIB/SEM. A cleaning pass was performed at 2kV to remove Ga implantation and surface damage, and final specimen thicknesses were around 100 nm. Element mapping was performed using energy-dispersive X-ray spectroscopy (EDS), measured in the TEM using a Super-X EDS system. Signal analysis was done using Bruker Esprit 1.9.

The films were electrically characterized on parallel plate capacitors with Pt top electrodes deposited through a shadow mask by direct-current sputtering. Devices were driven from the bottom Pt electrode and sensed from the top Pt contact at an excitation frequency of 10 kHz. The maximum electric field was gradually reduced for each nested loop for determining the coercive field. PE loops were generated with a Precision Multiferroic system from Radiant Technologies.

**Supporting Information**

Supporting Information is available from the Wiley Online Library or from the author.

**Acknowledgements**

This work was co-authored by Colorado School of Mines and the National Renewable Energy Laboratory, operated by the Alliance for Sustainable Energy, LLC, for the U.S. Department of Energy (DOE) under Contract No. DE-AC36-08GO28308. Funding was provided by the DARPA Tunable Ferroelectric Nitrides (TUFEN) program (DARPA-PA-19-04-03) as a part of Development and Exploration of FerroElectric Nitride Semiconductors (DEFENSE) project (diffraction, microscopy, and electrical characterization), and by Office of Science (SC), Office of Basic Energy Sciences (BES) as part of the Early Career Award "Kinetic Synthesis of Metastable Nitrides" (material synthesis and in-situ monitoring). The authors also express their appreciation to Dr. Wanlin Zhu and Prof. Susan Trolier-McKinstry of the Pennsylvania State University for providing Pt/TiO$_2$/SiO$_2$/Si substrates. We also thank Dr. Jeff Alleman for assistance with setting up the sputter system at NREL. The data affiliated with this study are available from the corresponding author upon reasonable request. The views expressed in the article do not necessarily represent the views of the DOE or the U.S. Government.

References

[1] R. Wang, X. Liu, I. Shih, Z. Mi *Appl. Phys. Lett*. **2015**, 106, 261104.

[2] W.-C. Hsu, D.-H. Huang, Y.-S. Lin, Y.-J. Chen, J.-C. Huang, C.-L. Wu *IEEE Trans. Electron Devices* **2006,** 53, 406.

[3] M. Akiyama, T. Kamohara, K. Kano, A. Teshigahara, Y. Takeuchi, N. Kawahara *Adv. Mater.* **2009**, 21, 593.

[4] K.R. Talley, R. Sherbondy, A. Zakutayev, and G.L. Brennecka *J. Vac. Sci. Technol., A* **2019**, 37, 060803.

[5] S. Fichtner, N. Wolff, F. Lofink, L. Kienle, and B. Wagner *J. Appl. Phys.* **2019**, 125, 114103.

[6] S. Yasuoka, T. Shimizu, A. Tateyama, M. Uehara, H. Yamada, M. Akiyama, Y. Hiranaga, Y. Cho, H. Funakubo *J. Appl. Phys.* **2020**, 128, 114103.

[7] S. Fichtner, T. Reimer, S. Chemnitz, F. Lofink, B. Wagner *APL Mater.* **2015**, 3, 116102.

[8] A. Zukauskaite, G. Wingqvist, J. Palisaitis, J. Jensen, P.O.Å. Persson, R. Matloub, P. Muralt, Y. Kim, J. Birch, L. Hultman *J. Appl. Phys.* **2012**, 111, 093527.

[9] M. Akiyama, K. Umeda, A. Honda, T. Nagase *Appl. Phys. Lett.* **2013**, 102, 021915.

[10] G. Wingqvist, F. Tasnádi, A. Zukauskaite, J. Birch, H. Arwin, L. Hultman *Appl. Phys. Lett.* **2010**, 97, 112902.

[11] Y. Lu, M. Reusch, N. Kurz, A. Ding, T. Christoph, M. Prescher, L. Kirste, O. Ambacher, A. Žukauskaitė *APL Mater*. **2018**, 6, 076105.

[12] M. Akiyama, K. Kano, A. Teshigahara *Appl. Phys. Lett*. **2009**, 95, 162107.

[13] O. Zywitzki, T. Modes, S. Barth, H. Bartzsch, P. Frach *Surf. Coat. Technol.* **2017**, 309, 417.

[14] A. Žukauskaitė, C. Tholander, F. Tasnádi, B. Alling, J. Palisaitis, J. Lu, P.O.Å. Persson, L. Hultman, J. Birch *Acta Mater*. **2015**, 94, 101.

[15] C.S. Sandu, F. Parsapour, S. Mertin, V. Pashchenko, R. Matloub, T. LaGrange, B. Heinz, P. Muralt *Phys. Status Solidi A* **2019**, 216, 1800569.

[16] F. Tasnádi, B. Alling, C. Höglund, G. Wingqvist, J. Birch, L. Hultman, I.A. Abrikosov *Phys. Rev. Lett*. **2010**, 104, 137601.

[17] K.R. Talley, S.L. Millican, J. Mangum, S. Siol, C.B. Musgrave, B. Gorman, A.M. Holder, A. Zakutayev, G.L. Brennecka *Phys. Rev. Mater.* **2018**, 2, 063802.

[18] S. Mishin, Y. Oshmyansky, presented at IEEE Int. Ultrason. Symp. Kobe, Japan, Oct., **2018**.

[19] A. Brudnik, A. Czapla, E. Kusior *Thin Solid Films* **2005**, 478, 67.

[20] D.L. Ma, H.Y. Liu, Q.Y. Deng, W.M. Yang, K. Silins, N. Huang, Y.X. Leng *Vacuum* **2019**, 160, 410.

[21] W. Wang, W. Yang, Z. Liu, H. Wang, L. Wen, G. Li *Sci. Rep.* **2015**, 5, 11480.

[22] S. Tungasmita, J. Birch, L. Hultman, E. Janzén, J.-E. Sundgren *Mater. Sci. Forum* **1998**, 264–268, 1225.

[23] P. Pigeat, T. Easwarakhanthan *Thin Solid Films* **2008**, 516, 3957.

[24] S.R. Kirkpatrick, S.L. Rohde, D.M. Mihut, M.L. Kurruppu, J.R. Swanson III, D. Thomson, J.A. Woollam *Thin Solid Films* **1998**, 332, 16.

[25] B. Saha, S. Saber, G.V. Naik, A. Boltasseva, E.A. Stach, E.P. Kvam, and T.D. Sands *Phys. Status Solidi B* **2015**, 252, 251.

[26] A. Bousquet, L. Spinelle, J. Cellier, and E. Tomasella *Plasma Processes Polym.* **2009**, 6, S605.

**Improving Reproducibility of Sputter Deposited Ferroelectric Wurtzite Al$_{0.6}$Sc$_{0.4}$N Films using In-situ Optical Emission Spectrometry**

*Daniel Drury\* , Keisuke Yazawa , Allison Mis, Kevin Talley, Andriy Zakutayev, Geoff L. Brennecka\**

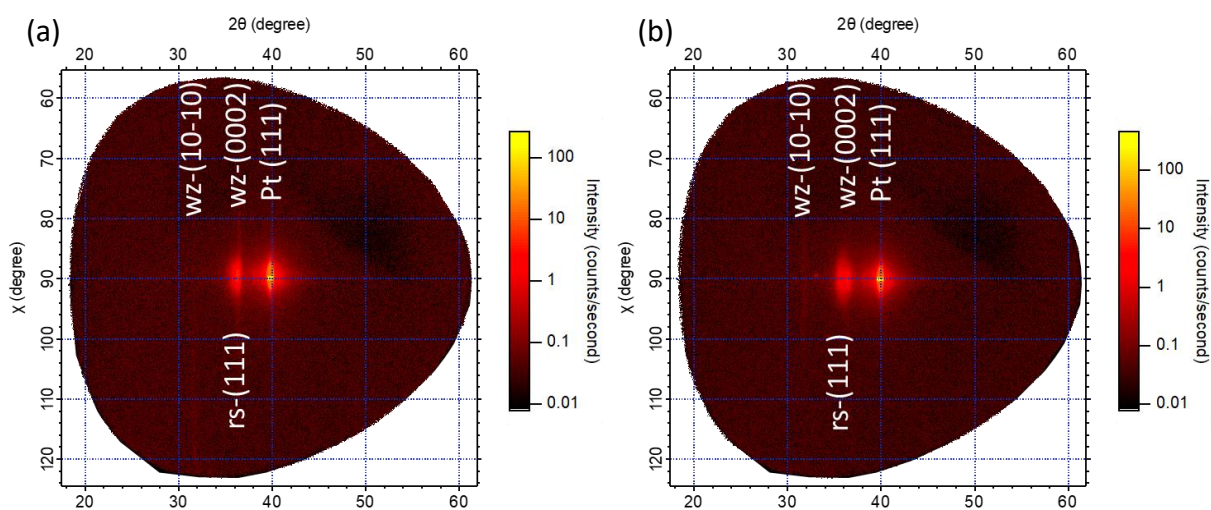

**Figure S1**. 2-dimensional XRD patterns of (a) wz-AlScN and (b) wz/rs-AlScN.

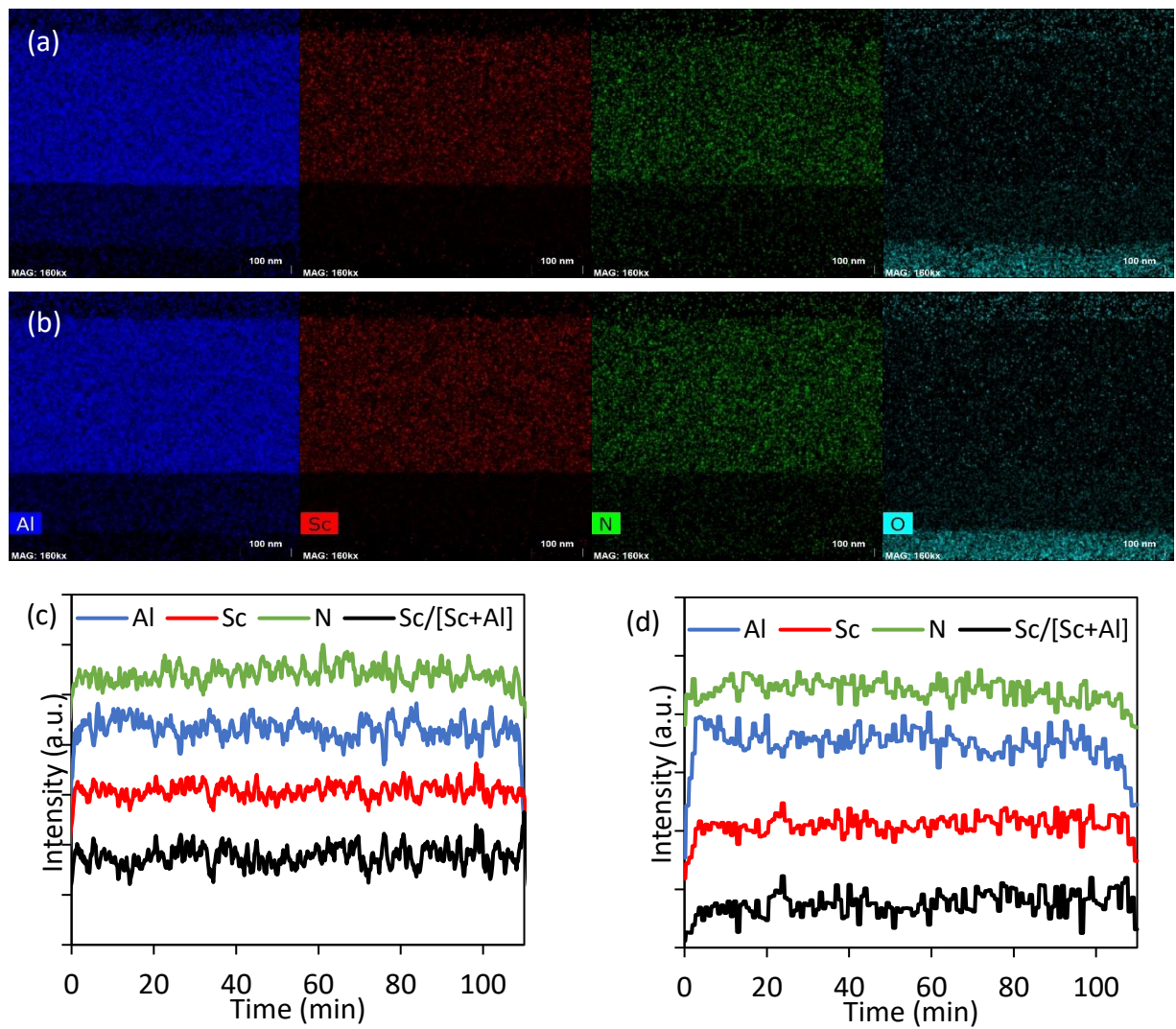

**Figure S2.** TEM-EDS maps showing Al, Sc, N, and O signal in $Al_{0.6}Sc_{0.4}N$ for (a) wz-AlScN sample and (b) wz/rs-AlScN. Stacked plot of the EDS profiles for (c) wz-AlScN and (d) wz/rs-AlScN where t=0 corresponds to the $Al_{0.6}Sc_{0.4}N$/Pt interface.

The emission lines from the excited states of molecular $N_2$(I) at 337.0 nm and atomic Ar(I) at 696.5 nm are presented in this work. The excited molecular $N_2$ line corresponds to a transition between the zero-point vibrational levels of two electronic states represented by $C^3\Pi_u \rightarrow B^3\Pi_g$ which needs 11.1 eV to occur. The excited Ar(I) state requires a 13.6 eV to occur when populated by the ground state of the neutral Ar. This makes Ar(I) at 696.5 nm a useful parameter to represent the state of electrons with moderate energy.[26]

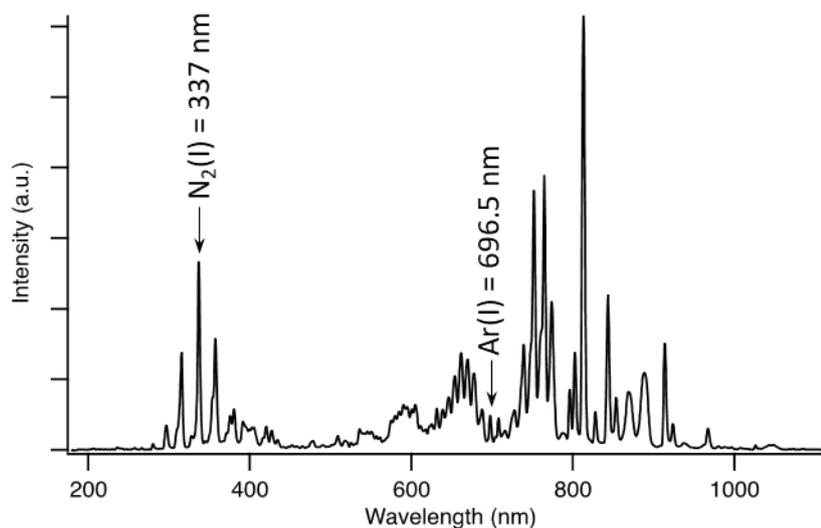

**Figure S3**. Glow discharge optical emssion spectrum with $N_2$(I) and Ar(I) lines marked.

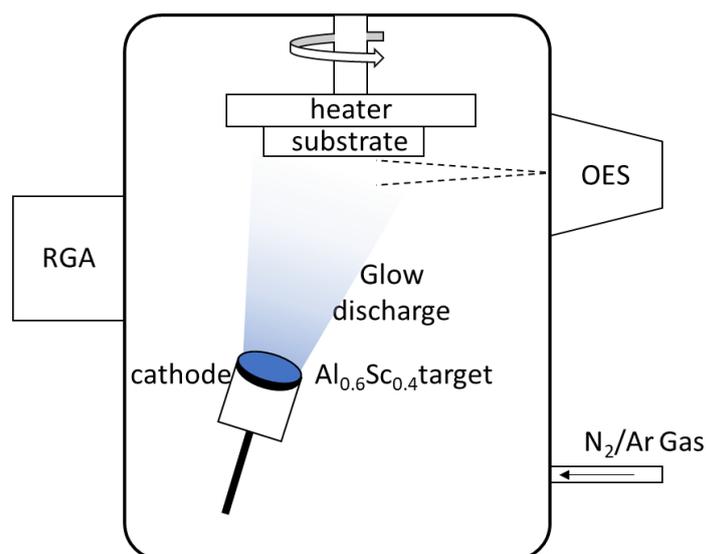

**Figure S4**. Sputter deposition chamber schematic depicting OES configuration.